# Solid Phase Recrystallization and Dopant Activation in Arsenic Ion-Implanted Silicon-On-Insulator by UV Laser Annealing

Toshiyuki Tabata[1], Fabien Rozé[1], Pablo Acosta Alba[2], Sébastien Halty[1], Pierre-Edouard Raynal[1], Imen Karmous[1], Sébastien Kerdilès[2], Fulvio Mazzamuto[1]

[1] Laser Systems & Solutions of Europe (LASSE), 145 rue des Caboeufs, 92230 Gennevilliers, France
[2] Université Grenoble Alpes, CEA-LETI, 17 rue des Martyrs, 38054 Grenoble, France
Phone: +33-1-4111-2720   E-mail: toshiyuki.tabata@screen-lasse.com

**Abstract**
UV laser annealing (UV-LA) enables surface-localized high-temperature thermal processing to form abrupt junctions in emerging monolithically stacked devices, where applicable thermal budget is restricted. In this work, UV-LA is performed to regrow a SOI layer partially amorphized by arsenic ion implantation and to activate the dopants. In a microsecond scale (~$10^{-6}$ s to ~$10^{-5}$ s) UV-LA process, monocrystalline solid phase recrystallization and dopant activation without junction deepening is evidenced, thus opening various applications in low thermal budget integration flows.

## 1. Introduction

Nowadays, to further explore alternative scaling paths, monolithically stacked devices are emerging [1-8]. However, vertical stacking of multiple functional layers brings severe limitation of the thermal budget applicable to top-tier devices (e.g., 500 °C for 2 h [3,6]), because bottom-tier ones must preserve their functions and performances during subsequent thermal processing steps. One of the most critical challenges is the formation of junctions on the top layers (e.g., source and drain [9] and their extension [8] for metal–oxide–semiconductor field-effect transistors (MOSFET), back-surface passivation for backside-illuminated complementary MOS imager sensors (BSI-CIS) [10,11]). If conventional annealing processes such as high temperature furnace and rapid thermal processing (RTP) are not compatible, is generally used a lower temperature process called solid phase recrystallization (SPR), which can be performed at 500 to 600 °C in ion-implanted silicon (Si) [12,13]. However, such low-temperature SPR undergoes significant reduction of recrystallization rate [14]. To compensate it, the annealing time must be extended, but it may result in deactivation of dopants. Ion implantation introduces a high concentration of interstitials (i.e., point defects) underneath the amorphized layer, and a thermal processing transforms them into dislocation loops (i.e., extended defects) [15]. These so-called end-of-range (EOR) defects newly release Si interstitials towards the neighboring surface and interface to reduce the free energy of system [16,17]. Those Si interstitials may interact with substitutional dopants and lead to their deactivation [17].

UV laser annealing (UV-LA) can be advantageous for SPR because it allows to reach a high temperature while conserving the functionality of surrounding devices thanks to its short timescale and shallow irradiation absorption. In our previous work, a UV-LA SPR process has been demonstrated on a 22-nm-thick silicon-on-insulator (SOI) substrate partially amorphized by ion implantation [18,19]. Then, a maximum crystallization rate of 1.8 nm/sequence was obtained for an effective irradiation time ($t_{eff}$) of ~$10^{-7}$ s, and thermally independent multiple irradiations were necessary to recrystallize the entire amorphized layer. Considering a typical amorphization thickness in fully depleted SOI devices (e.g., 5 to 7 nm for the extension [8], 13 to 20 nm for the source and drain [1,17]), this crystallization rate is acceptable. However, other devices such as power IC may require recrystallization of a much thicker layer (e.g., 150 nm for a SOI-based lateral double-diffused MOSFET [20]), and a more effective control of recrystallization, particularly extending $t_{eff}$, may be mandatory to guarantee a reasonable productivity for high volume manufacturing.

In this work, we present a UV-LA SPR process with a $t_{eff}$ of ~$10^{-6}$ s to ~$10^{-5}$ s performed on a relatively thick SOI structure amorphized by arsenic (As) ion implantation for roughly a half of its thickness. Sheet resistance, surface morphology, atomic diffusion, As activation, and SOI microstructure after UV-LA are discussed, evidencing a SPR completion among applied process conditions.

## 2. Experimental

A 70-nm-thick SOI wafer having a 145-nm-thick buried oxide (BOX) was used as starting material. The wafer had a (100) surface. Arsenic ion implantation was performed at room temperature (RT) at 19 keV with a total dose of 4 × $10^{15}$ cm$^{-2}$. A 37-nm-thick amorphization was confirmed by cross-sectional transmission electron microscopy (TEM). The wafer was then submitted to UV-LA at RT in air, varying both laser fluence (LF) and $t_{eff}$ in only one laser irradiation sequence to control the heat generated in the SOI structure. Sheet resistance ($R_{sq}$) was measured by a standard four-point probe. Surface morphology was observed by atomic force microscopy (AFM). Arsenic and oxygen (O) diffusion within the SOI layer were measured by secondary ion mass spectroscopy (SIMS). Crystallinity after UV-LA was discussed by cross-sectional TEM. Active shallow donor and inactive defect-related deep donor/acceptor concentrations were estimated by an analytical model developed for electrochemical capacitance voltage profiling (ECVP) [21].

## 3. Results and Discussion

Figure 1 shows the $R_{sq}$ plots obtained with two different LF values ($LF_1 < LF_2$) as a function of $t_{eff}$. By increasing $t_{eff}$, the $R_{sq}$ value monotonically decreased, meaning that the amorphized film is progressively re-crystallized and the As atoms are activated in the SOI layer. Both LFs show a similar tendency, having a plateau at around 100 ohm/sq, where we suppose that a SPR is completed over the amorphized region. A smaller LF required a longer $t_{eff}$ to reach this plateau. The beginning point of the plateau for each LF is indicated by an arrow. The $R_{sq}$ value further decreased at the longest $t_{eff}$ for $LF_2$, suggesting that the ion-implanted SOI is melted and that the As atoms diffuse thereby.

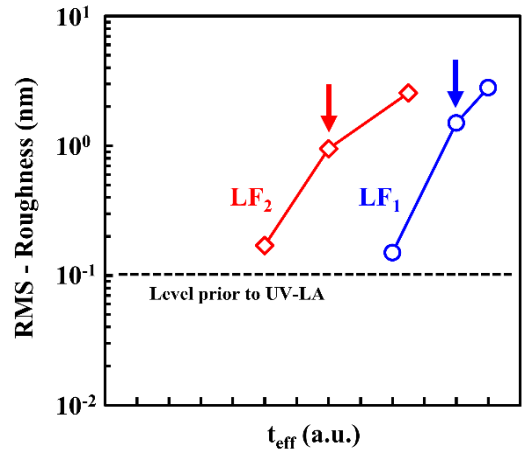

Fig. 2 Root-mean-square (RMS) values of surface roughness measured by AFM with a 30 × 30 μm² scan before and after UV-LA. The scale of X axis is aligned with that of Fig. 1 and the beginning point of the plateau for each LF is indicated by an arrow.

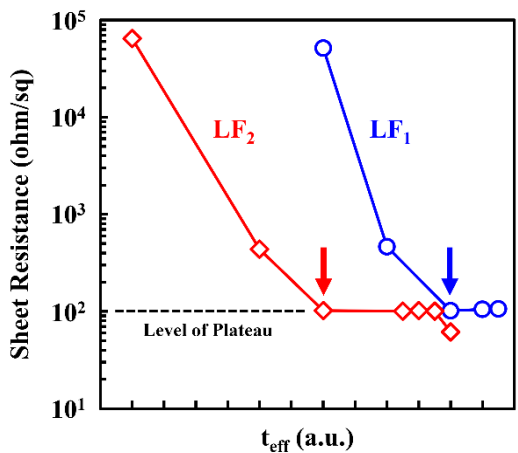

Fig. 1 Sheet resistance ($R_{sq}$) plots with different LF and $t_{eff}$. The beginning point of the plateau for each LF is indicated by an arrow.

Figure 2 shows the plots of root-mean-square (RMS) values of surface roughness measured by AFM with a 30 × 30 μm² scan. The scale of X axis is aligned with that of Fig. 1 and the beginning of the plateau for each FL is indicated by an arrow as well. Before reaching the $R_{sq}$ plateau, only a small RMS value increase was observed. At the beginning of the $R_{sq}$ plateau, however, the RMS value became much more pronounced and was continuously increased with $t_{eff}$ increase. It should be noted that the irradiated laser beam was not perfectly optimized and probably still underwent a non-uniform shape. If so, one may speculate that a heat gradient is introduced in the SOI layer plane, resulting in a local variation of recrystallization rate. This is a possible reason of increasing the RMS value of surface roughness during UV-LA SPR. Otherwise, if the heating continues even after a SPR completion, surface roughness may further degrade because of the recombination of Si interstitials, whose sources are either of the ion implantation damage or the subsequently formed EOR defects, proceeding at the SOI surface and the interface with the BOX layer [16,17]. However, as our $t_{eff}$ range makes negligible a Si self-diffusion length calculated at the melting temperature of amorphous Si (1420 K [22]) in crystalline Si (~$10^{-3}$ to ~$10^{-2}$ nm [23]), the Si interstitials cannot travel towards the SOI surface.

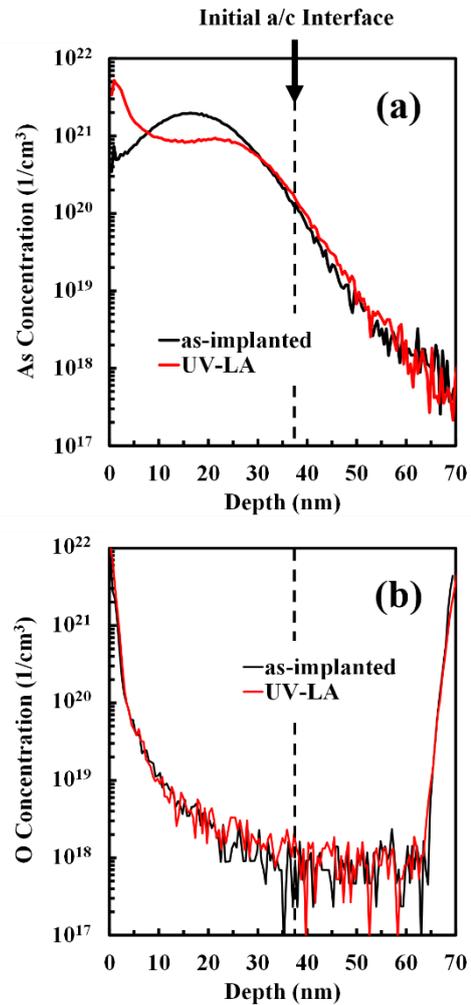

Fig. 3 SIMS profiles of (a) As and (b) O atoms before and after UV-LA with the $LF_1$ and $t_{eff}$ giving the beginning of the $R_{sq}$ plateau. The initial amorphous/crystal (a/c) interface in the SOI layer is also indicated by an arrow and a dotted line.

Figure 3 shows the As and O SIMS profiles before and after UV-LA at the $LF_1$ and $t_{eff}$ giving the beginning of the $R_{sq}$ plateau. An important evidence of SPR is that there is no O diffusion at this condition (Fig. 3(b)). When melting of an ion-implanted Si happens, O incorporation is observed even in a nitrogen ambient for a much shorter $t_{eff}$ (~$10^{-7}$ s) [24]. On the other hand, a clear As segregation towards the surface was observed in this SPR condition in the initially amorphized layer (Fig. 3(a)), known as "snowplow effect" [25]. This characteristic As redistribution can be beneficial for lowering contact resistivity of transistors [26,27].

Figure 4 shows dark-field cross-sectional TEM images taken before and after UV-LA at the $LF_1$ and $t_{eff}$ giving the beginning of the $R_{sq}$ plateau. A partial amorphization of the SOI layer by the As ion implantation is clear and the amorphized thickness can be estimated to be about 37 nm (Fig. 4(a)). After the UV-LA process, the amorphized part is fully recrystallized (Fig. 4(b)). Although the applied magnification is not high enough to observe crystal lattice patterns, the overall black-and-white contrast is uniform in the whole regrown SOI layer, and no clear grain boundary can be found. This observation confirms a SPR completion at this UV-LA condition. Nevertheless, some change of the contrast, possibly related to local defects, seems observed. As already discussed, the Si interstitials cannot reach the SOI surface in our UV-LA process timescale. Instead, the solid solubility limit of As in Si (~$3 \times 10^{20}$ at 1100 °C [28]) and the As surface segregation (Fig. 3(a)) suggest precipitation of quite many As atoms near the SOI surface. Again, in the given $t_{eff}$, are allowed neither the Si interstitials travelling towards the SOI layer bottom nor the EOR defect evolution. Therefore, the contrast change suspected between the initial a/c interface and the SOI layer bottom might be related to the ion implantation damage left in the EOR region and not fully annihilated in our process. Anyhow, it needs to be confirmed by a more adequate TEM configuration.

About the crystallinity after short timescale SPR, some interesting cases to compare can be found in the literature. In the case of a shorter $t_{eff}$ (~$10^{-7}$ s) from our previous work [18, 19], the calculated Si self-diffusion length in crystalline Si is further reduced (~$10^{-4}$ to ~$10^{-3}$ nm [23]). Although the same situation can be expected, after a SPR completion in a phosphorus (P) (4 keV, $1 \times 10^{15}$ cm$^{-2}$) [19] or As (9 keV, $1 \times 10^{15}$ cm$^{-2}$) [18] doped 22-nm-thick SOI layer, no obvious defect was pointed out in cross-sectional TEM of a higher magnification than Fig. 4 of this work (note, however, that defects may be invisible in TEM depending on their geometric relations [29]). Since the as-implanted peak concentration of dopants (~$5 \times 10^{20}$ cm$^{-3}$ for the applied P condition [30] and ~$1 \times 10^{21}$ cm$^{-3}$ for the As one [31]) is lower in these two cases than in this work (~$2 \times 10^{21}$ cm$^{-3}$), the precipitation of dopants may be greatly suppressed. This means that the defectivity possibly left in our SPR process may be reduced by optimizing the ion implantation condition. In addition, it is known that the point defects coalesce into intermediate defect configurations (IDCs) and can be observed in TEM (a typical size is ~2 nm) [15]. These IDCs are observed at 400 to 700 °C and dissolve at a higher temperature [15]. Although a kinetic, rather than thermodynamic, feature should be considered in our UV-LA, a possible difference of reached maximal temperature might also have an impact on the defectivity. Furthermore, contrary to this work, a flat surface (a RMS value of ~0.12 nm) was maintained probably because of a better laser uniformity [19]. In the case of a longer $t_{eff}$ (~$10^{-3}$ to ~$10^{-2}$ s) [32,33], the calculated Si self-diffusion length in crystalline Si reaches ~$10^{-2}$ to ~$10^{-1}$ nm [23]. Considering the typical Si-Si bonding length in crystalline Si (~0.23 nm [34]), the EOR defects may grow in this timescale. Indeed, the EOR defect evolution is then confirmed by weak-beam dark-field TEM.

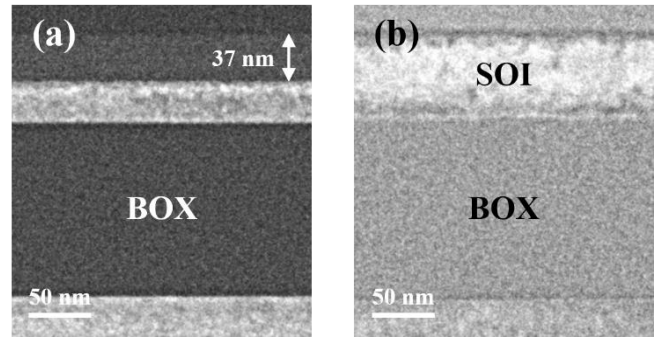

Fig. 4 Dark-field cross-sectional TEM images taken (a) before (i.e., as-implanted) and (b) after UV-LA with the $LF_1$ and $t_{eff}$ giving the beginning of the $R_{sq}$ plateau.

At the end, to study As activation after the UV-LA induced SPR, ECVP was performed on the sample annealed at $LF_1$ for the $t_{eff}$ giving the beginning of the $R_{sq}$ plateau. In a typical ECVP measurement, a doped semiconductor surface is in contact with an electrolyte and forms a Schottky junction. Applying a reverse polarization field to the junction, the doped semiconductor is depleted, and a space charge zone is formed. The measured capacitance ($C$) is written as the following [21]:

$$\frac{1}{C^2} = \frac{2(V_{bi} - V)}{\varepsilon_r \varepsilon_0 q N^* A^2} \qquad (1)$$

where $V_{bi}$ stands for the flat-band polarization field (called "built-in potential"), $V$ for the applied polarization field, $\varepsilon_r$ and $\varepsilon_0$ for the relative and vacuum dielectric constants, $q$ for the electron charge, $N^*$ for the concentration of ionized impurities, and $A$ for the junction area. If any deep levels exist, the $1/C^2$ versus $V$ plot shows a curvature. A fitting approach is therefore taken for all $1/C^2$ versus $V$ curves measured at different depths after cyclic etching, involving the active shallow donor and inactive defect-related deep donor/acceptor concentrations into $N^*$. The detailed procedure is explained elsewhere [21].

Figure 5 shows the depth profile of the shallow donor ($N_{sd}$) and deep donor/acceptor ($N_{dd} + N_{da}$) concentrations extracted by the $1/C^2$ versus $V$ curve fitting.

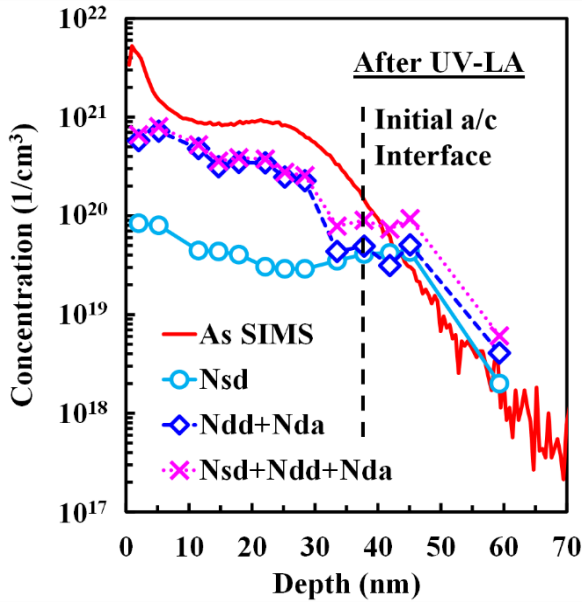

Fig. 5 Depth profiles of the shallow donor ($N_{sd}$) and deep donor/acceptor ($N_{dd} + N_{da}$) concentrations extracted by the $1/C^2$ versus $V$ curve fitting in ECVP for the sample annealed with the LF$_1$ and $t_{eff}$ giving the beginning of the R$_{sq}$ plateau. The initial amorphous/crystal (a/c) interface in the SOI layer is also indicated by a dotted line.

First, in the regions regrown by SPR and just below the initial amorphous/crystal (a/c) interface, considering a known range of the equillibrium solid solubility limit of As in Si (~1 × 10$^{20}$ to ~3 × 10$^{20}$ cm$^{-3}$ at 700 to 1100 °C [28], and note that the maximum temperature reached in our process is unknown), many substitutional As atoms seems deactivated and to form deep-level defects. As already discussed, no Si interstitials travels from the EOR region during our UV-LA SPR. Therefore, another deactivation pathway needs to be considered. In fact, a high As dose makes it happen [35]. To begin, an As$_n$ cluster is formed among substitutional As atoms. Coupling of this cluster with a vacancy (V$_{Si}$) happens (i.e., formation of As$_n$V), emitting an interstitial Si (I$_{Si}$), especially for n = 3, 4. This emitted I$_{Si}$ replaces a nearest substitutional As atom (i.e., release of an interstitial As (I$_{As}$)). The released I$_{As}$ makes As$_n$V grow (i.e., formation of As$_{n+1}$V). This process self-continues with time and stops with As$_4$V. Indeed, As$_n$V is supposed to introduce deep donor/acceptor levels in the Si bandgap [36,37]. These As$_n$V type defects might be observed as local variation of the contrast in the cross-sectional TEM image. In addition, it should be noted that there is a discrepancy between the As SIMS and $N_{sd} + N_{dd} + N_{da}$ profiles. This might represent either the mean number of charges trapped by the As$_n$V defects or the precipitation of As atoms in the Si lattice.

Second, in the region close to the SOI bottom, all As atoms are active. On the other hand, some deep donors/accepters still exist. Most of them are probably the point defects, which originally come from the ion implantation damage and are not fully annihilated in our process. Indeed, V$_{Si}$ is supposed to form deep-level defects as well [36].

## 4. Conclusions

We have investigated the UV-LA induced SPR on a partially amorphized 70-nm-thick SOI substrate to demonstrate a potential of stretching $t_{eff}$ up to a range of ~10$^{-6}$ s to ~10$^{-5}$ s (compared to ~10$^{-7}$ s of our previous work [18,19]) while maintaining a flat surface morphology, good crystal quality and high dopant activation level. The results have evidenced a single-crystal regrowth of the amorphized region by SPR. However, there remains some concerns to address in a future work.

First, the surface roughness was degraded during our UV-LA SPR. So far, the root cause seems a possible laser non uniformity. Second, the precipitation of As atoms was suspected over the SOI layer, especially near the surface where the As segregation was observed. Such excess As concentration was also supposed to trigger the deactivation of substitutional As atoms. Therefore, the initial As dose must be carefully optimized. In the case of a zero-second rapid thermal annealing followed by fast cooling (250 °C/s), the maximum carrier (i.e., substitutional As) concentration monotonically increases with increase of the annealing temperature [38]. This suggests maintaining the annealing temperature close to the upper limit (1420 K) during UV-LA SPR. Third, although the EOR defects seem hardly formed due to the negligible Si self-diffusion length, some deep levels were electrically still measured in the EOR region. This suggests suppressing the ion implantation damage to further improve the crystallinity of the EOR region after UV-LA SPR. Cryogenic ion implantation may help it [15,39,40].

**Acknowledgements**

The work realized by LASSE for this publication was supported by the IT2 project. This project has received funding from the ECSEL Joint Undertaking (JU) under grant agreement No 875999. The JU receives support from the European Union's Horizon 2020 research and innovation programme and Netherlands, Belgium, Germany, France, Austria, Hungary, United Kingdom, Romania, Israel.

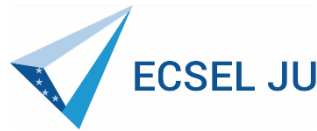
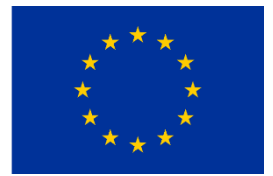